\documentclass[floatfix,twocolumn,showpacs,preprintnumbers,amsmath,amssymb,prl]{revtex4}
\newcommand{\etal}{{\it et al.}}

\usepackage{graphicx}
\usepackage{epstopdf}
\usepackage{dcolumn}
\usepackage{rotating}
\usepackage{setspace} 

\begin{document}

\title{Enhancing the Thermal Stability of Majorana Fermions with Redundancy Using Dipoles in Optical Lattices}

\author{Fei Lin and V.W. Scarola}
\affiliation{Department of Physics, Virginia Tech, Blacksburg, Virginia 24061 USA}

\date{\today}

\begin{abstract}
Pairing between spinless fermions can generate Majorana fermion excitations that exhibit intriguing properties arising from non-local correlations.  But simple models indicate that non-local correlation between Majorana fermions becomes unstable at non-zero temperatures.  We address this issue by showing that anisotropic interactions between dipolar fermions in optical lattices can be used to significantly enhance thermal stability.  We construct a model of oriented dipolar fermions in a square optical lattice.  We find that domains established by strong interactions exhibit enhanced correlation between Majorana fermions over large distances and long times even at finite temperatures, suitable for stable redundancy encoding of quantum information.  Our approach can be generalized to a variety of configurations and other systems, such as quantum wire arrays.    
\end{abstract}


\maketitle

\noindent
{\it Introduction:} The wide variety of optical lattice geometries offer unprecedented tunability in manipulating quantum degenerate gases into complex quantum states \cite{greiner:2002}.  Recent developments in the cooling of molecules (e.g., $^{40}\text{K} ^{87}\text{Rb}$) \cite{ospelkaus:2008} and magnetic atoms (e.g., $^{161}$Dy) \cite{lu:2012} imply that anisotropy in dipolar interactions will soon provide further opportunity to explore some of the most elusive yet compelling quantum states, entangled Majorana fermions (MFs).

Seminal lattice models demonstrate particle-like excitations that behave as MFs thanks to non-local symmetries \cite{kitaev:2001,kitaev:2003}.  They entangle with each other over large distances through string operator (SO) correlations.  In simple models SOs have straightforward definitions, e.g., fermion parity \cite{kitaev:2001}, with non-trivial consequences.  They signal underlying topological order with fascinating properties that have motivated proposals for topologically protected qubits \cite{kitaev:2003,nayak:2008}. The crossing of SOs is responsible for unusual anyonic braid statistics \cite{kitaev:2003,nussinov:2009}.  And SOs connecting these excitations also underlie theories of quantum state teleportation \cite{semenoff:2007,tewari:2008}.  

The zero-temperature properties of models hosting topological order set the stage for work connected to experiments.  Kitaev's two-dimensional (2D) Toric Code Hamiltonian \cite{kitaev:2003} motivated early proposals in optical lattices \cite{duan:2003, micheli:2006, weimer:2010}.  But the 1D Kitaev chain model \cite{kitaev:2001} is one of the simplest models supporting MF excitations.  Anticipation of non-local MF properties in 1D led to experimental proposals and experiments in both optical lattices \cite{jiang:2011,diehl:2011, kraus:2012} and solids \cite{kitaev:2001,lutchyn:2010, mourik:2012}.  But prospects for observing non-local correlation of MF pairs over long times and distances hinge on the stability of SOs \cite{nussinov:2009,chesi:2010}. 

SOs in important lattice models are unstable at non-zero temperatures. For example, SOs in the 2D Toric Code model vanish at long times and distances because of thermal excitations \cite{nussinov:2009, chesi:2010, castelnovo:2007, hastings:2011}.  Recent work also argues that MFs in lattice models of topological $p$-wave superconductors are sensitive to thermal fluctuations \cite{cheng:2010,bauer:2012}.  A general theorem \cite{hastings:2011} sets strict criteria for non-local correlations to remain resilient against thermal fluctuations.  Fortunately, recent calculations indicate that topological phases can be enhanced through: disorder \cite{wootton:2011}, and proximity coupling \cite{fidkowski:2011,fu:2008} to a reservoir in topological superconducting wires \cite{lutchyn:2010}. There are also proposals to go beyond 1D wires to multi-channel or 2D MF arrays \cite{potter:2010}.

\begin{figure}[t]
\includegraphics[clip,width=80mm]{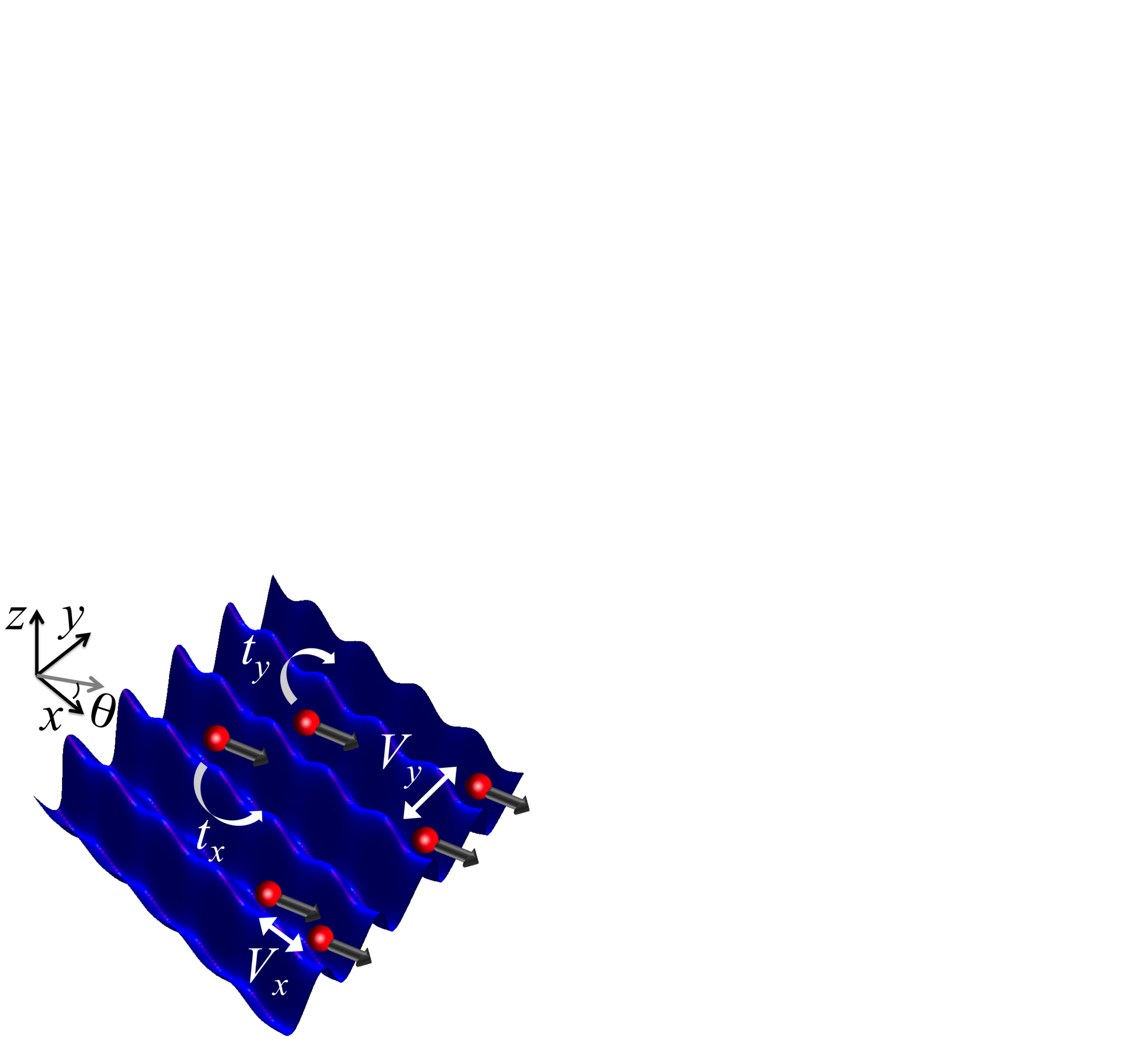}
\caption{(Color online) Schematic of dipolar fermions (spheres) in a 2D optical lattice.  Dipolar moments $\vec{p}$ (arrows on each sphere) align along an applied field, at an angle $\theta$ with the $x$-axis.}
\label{dipole_lattice}
\end{figure}

We propose that dipolar interactions in optical lattices \cite{baranov:2012} offer a powerful tool to stabilize the SOs in MF models.  We show that anisotropy in both the lattice and dipolar interactions electrostatically copy SOs to force excitations to form arrays of strings which we call {\it domains} in this work.  We thus propose a robust mechanism, the formation of domains with redundant MF edges, as a route to stabilize MFs, akin to quantum error correction schemes using redundant qubits \cite{shor:1995}.  We pair two methods [quantum Monte Carlo (QMC) and mean field theory] to solve a  model of dipolar fermions to demonstrate that domain formation in electrostatically coupled Kitaev chains significantly enhances the stability of SOs.  QMC here is unbiased and shows the thermal stability of domains while our mean field theory (which agrees with QMC within regimes of applicability) explicitly reveals MFs. 

\noindent
{\it Model:}  We first consider a Hubbard model of dipolar fermions in an $L\times L$ optical lattice and then discuss a specific parameter regime. In Fig.~\ref{dipole_lattice} fermions with dipolar moment $\vec{p}$ can hop between nearest neighbor (NN) sites.  A large optical lattice depth along the $y$ direction strongly suppresses hopping in the $y$ direction.  $V_x(\theta)=D^2(1-3\cos^2\theta)/r_0^3$ ($V_y=D^2/r_0^3$) is the $x$ ($y$) component of the NN dipolar fermion interaction. Here $D^2\sim \vec{p}^2$ and $r_0$ is lattice constant. We can tune $\theta$ so that the NN dipolar interaction is attractive along the $x$-direction. We construct a Hubbard model capturing the above features:
\begin{eqnarray}
& &H_D=-\sum_{i,j}\left(t_{x}a_{i,j}^{\dagger}a_{i+1,j}^{\phantom{\dagger}}+t_{y}a_{i,j}^{\dagger}a_{i,j+1}^{\phantom{\dagger}}+h.c.\right)\nonumber\\
&+&\sum_{i,j}\left[V_x(\theta)n_{i,j}n_{i+1,j}+V_y n_{i,j}n_{i,j+1}-\mu_{0}n_{i,j}\right],
\label{dipolarH}
\end{eqnarray}
where we have open(periodic) boundary condition in the $x(y$) directions.  $a_{i,j}^{\dagger}$ creates a spinless fermion at the site $(i,j)$ and $n_{i,j}=a_{i,j}^{\dagger}a_{i,j}^{\phantom{\dagger}}$. $t_{x}(t_{y})$ is the hopping energy between NN sites in the $x(y)$ direction.  $\mu_{0}$ is the chemical potential.

For a range of $\theta$ yielding $V_{x}<0$ the ground state of Eq.~(\ref{dipolarH}) is stable and exhibits $p$-wave pairing.  For $t_{x}=t_{y}$ functional renormalization group \cite{bhongale:2012} and mean field theory \cite{liu:2012} calculations show a BCS paired state for long-range dipolar interactions consistent with short-range interactions in Eq.~(\ref{dipolarH}) \cite{cheng:2010}.  $p$-wave pairing between neighbors along $x$-rows can be modeled by real-space attraction: $\exp(\bold{i}\Phi_{i,j})\vert  \Delta  \vert a_{i+1,j}^{\dagger}a_{i,j}^{\dagger} + h.c.$, where $\Phi_{i,j}$ and $\vert \Delta \vert$ are the phase and magnitude of the pairing field within an $x$-row.  But for $t_{y} \ll t_{x}$ the system can be analyzed with Luttinger liquid theory to show that weakly coupled 1D dipolar systems also posses $p$-wave pairing order with algebraically decaying pairing correlations \cite{huang:2009}.  For $t_y\ll \vert \Delta \vert$, Josephson tunneling between paired states  contributes an energy: $\sim -t_{y} ^{2} \cos( \Phi_{i,j}-\Phi_{i,j+1})$, which aligns the phase of the pairing field between each $x$-row, $\Phi_{i,j}-\Phi_{i,j+1}\rightarrow 0$. Hereafter, we assume a uniform pairing field to motivate a thermally stable MF model.  Increasing $t_y$ should adiabatically connect the coupled-1D \cite{huang:2009} and 2D square lattice limits \cite{bhongale:2012,liu:2012}.

\noindent
{\it Effective Model:}  We perform a mean field decoupling of the attractive dipolar interaction term in Eq.~(\ref{dipolarH}) to establish the centerpiece of our study \cite{supplementarymaterial}:
\begin{equation}
H_F=\sum_j H_K^j+V_y\sum_{i,j}\left (n_{i,j}-\frac{1}{2} \right ) \left (n_{i,j+1}-\frac{1}{2}\right ),
\label{fermionH}
\end{equation}
where the Hamiltonian for the $j$th Kitaev chain is $H_K^j=-t\sum_{i}\left (a_{i,j}^{\dagger}-a_{i,j}^{\phantom{\dagger}}\right) \left (a_{i+1,j}^{\dagger}+a_{i+1,j}^{\phantom{\dagger}}\right )-\mu n_{i,j}$. 
At the Hartree-Fock level the chemical potential renormalizes to $\mu=\mu_{0}+2\langle n_{i,j}\rangle |V_x(\theta)|-V_y/2$ and the hopping becomes 
$t=t_{x}-  |V_x(\theta)|  \langle a_{i+1,j}^{\dagger}a_{i,j}^{\vphantom{\dagger}} \rangle$, which is our energy unit.  In Eq.~(\ref{fermionH}), we tuned $V_{x}$ to match the pairing term with the renormalized hopping by setting $t_{x}=|V_x(\theta)|\langle a_{i+1,j}^{\dagger}a_{i,j}^{\dagger}+ a_{i+1,j}^{\dagger}a_{i,j}^{\vphantom{\dagger}}\rangle$. MFs can arise away from this particular point, which is guaranteed by the presence of a gap in the energy spectrum of $H_{F}$ \cite{dorier:2005}.  $t_y$ is energetically negligible but is included as a second order
effect by setting $\Phi_{i,j}=0$.   We work near half filling $\langle n \rangle=1/2$, i.e., $\mu=0$.

Eq.~(\ref{fermionH}) describes an array of strongly interacting Kitaev chains, whose ground state is $2^L$-fold degenerate \cite{supplementarymaterial}, which is not explicit in Eq.~(\ref{dipolarH}). Our direct QMC simulations on Eq.~(\ref{dipolarH}) show the emergence of precisely the same set of degeneracies expected from Eq.~(\ref{fermionH}) for the parameters given by the Hartree-Fock decoupling \cite{supplementarymaterial,lin:2013}.

\begin{figure}[t]
\begin{center}
\vspace{-.6in}
\includegraphics[width=4in]{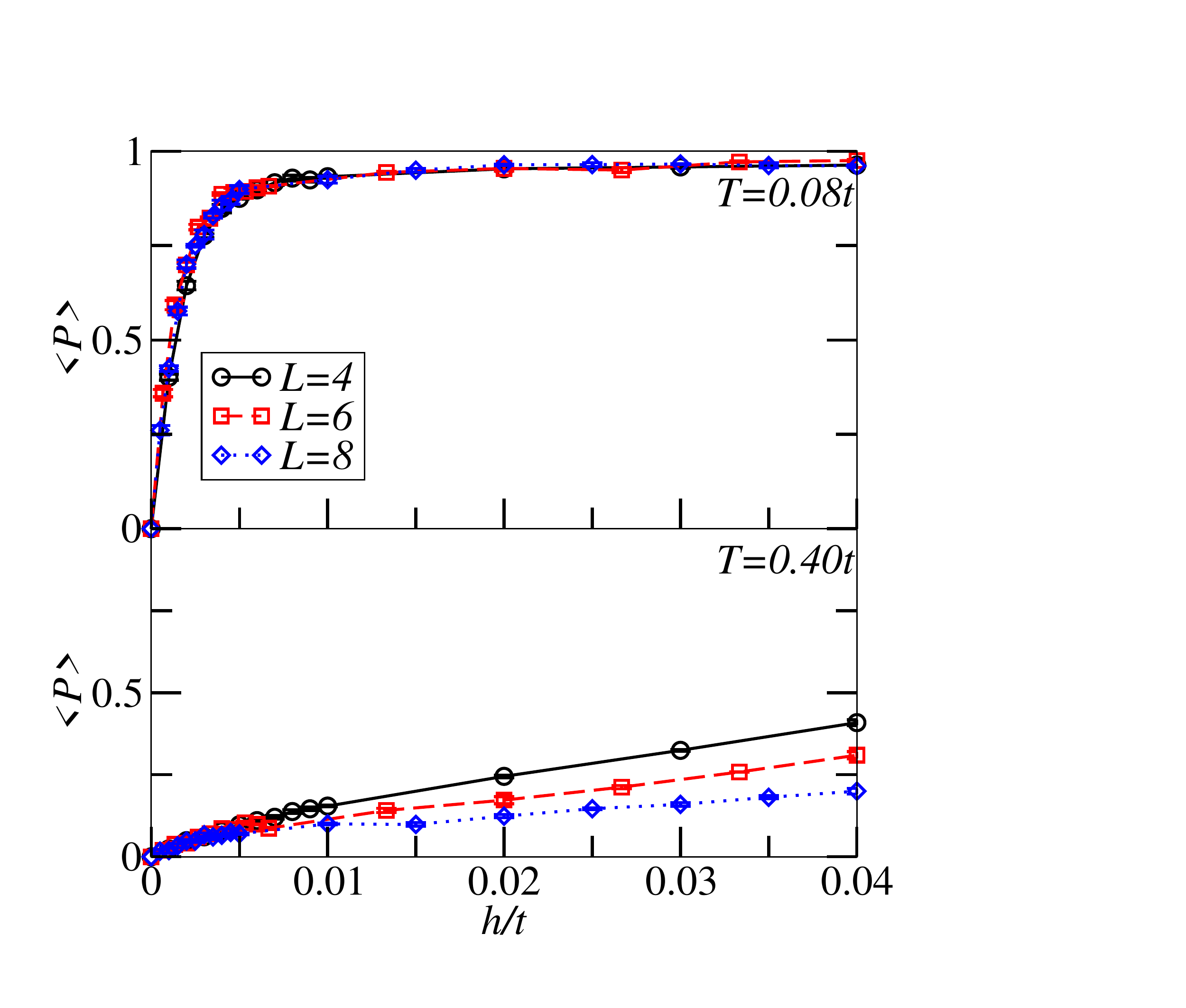}
\vspace{-2.8in}

\includegraphics[width=1.5in]{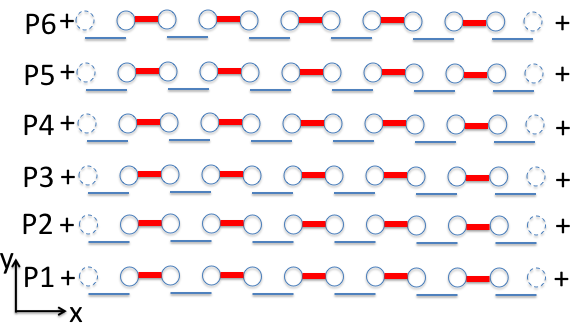}
\hspace{-0.8in}
\vspace{0.17in}

\includegraphics[width=1.4in]{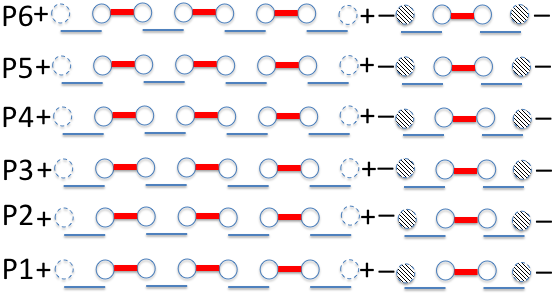}
\hspace{0.72in}
\vspace{0.4in}

\end{center}
\caption{(Color online) The thermal expectation value of SOs from QMC as a function of an applied global field for several system sizes for $V_y=4.8t$ and $\mu=0$.  
The top (bottom) panel shows data for a characteristic low (high) temperature.   The insets show schematic examples of a MF domain that breaks up into two MF domains at high temperatures. ``+'' in the figures is fermion parity for the entire chain, and each chain has the same parity for the one configuration drawn. Empty dashed circles denote empty MF edge states; hatched circles denote MF edge states occupied by one particle per row.}
\label{PvsZ2H}
\end{figure}

\noindent
{\it Mechanism for Stabilizing MFs:}  Eq.~(\ref{fermionH}) is a highly non-trivial many-body model.  It maps onto an intractable quantum spin compass model \cite{supplementarymaterial,dorier:2005}.  Below we argue that the inter-chain interactions stabilize correlation between edge $y$-columns of MFs.  

We use mean field theory to show that Eq.~(\ref{fermionH}) reduces to a MF model \cite{supplementarymaterial}.  Consider a pair of MF operators, $c_{2i,j}$ and $c_{2i-1,j}$, for each site of the lattice, $(i,j)$, where $a_{i,j}^{\dagger}=(c_{2i-1,j}-\bold{i}c_{2i,j})/2$ \cite{kitaev:2001}.  We impose a mean field decoupling of the $V_y$ term, using a 2-site unit cell along the $y$ direction.  Each site of the unit cell corresponds to sublattice A or B. We thus have $H_M^{\alpha}=\bold{i}t\sum_{i}c_{2i,\alpha}c_{2i+1,\alpha} +(\bold{i}\tilde{\mu}_{\alpha}/2)\sum_{i}c_{2i-1,\alpha}c_{2i,\alpha}$, where $\alpha \in \{ A,B\}$ denotes sublattice and the renormalized chemical potential, $\tilde{\mu}_{\alpha}=\mu+\bold{i}V_y\langle c_{2i-1,\alpha}c_{2i,\alpha} \rangle$. Furthermore, we can show \cite{supplementarymaterial} that the ground state avoids strong $V_{y}$ by setting $\langle c_{2i-1,\alpha}c_{2i,\alpha} \rangle=0$ for $V_y>4t$.  This leads to two columns of localized MF states, one at each edge.  

Solutions of $H_M$ exhibit domains with MF edge states along $y$-columns (Fig.~\ref{PvsZ2H}) \cite{supplementarymaterial}. 
Note that the $V_{y}$ term in Eq.~(\ref{fermionH}) leads to a chemical potential staggered along $y$ columns, which binds MFs along $y$ 
but leaves them to propagate along $x$. An energy penalty,  $\sim V_{y}$, will result if only one row changes its parity.  
The inter-row interaction therefore increases the dimension of the MF edge state (from a point particle to a $y$-column) to establish the mechanism for enhancing the stability of the non-local MF state against thermal fluctuations. The entire ground state can thus be regarded as a redundantly encoded qubit of several MFs.  Along these lines, mean field theory 
suggests the following Gutzwiller projected wave function:
$
\prod_{i,j=1}^{L}\left(1-n_{i,j}n_{i,j+1}\right)\phi^j_{\rm BCS},
$
where $\phi^j_{\rm BCS}$ is the BCS wave function hosting MFs in the $j$th $x$-row.  

Thermally stable non-local correlation implies that $y$-columns of MF pairs at $i=1$ and $i=L$ host real dipoles in a superposition that remains robust against thermal excitations.  To establish robustness we note that the Hilbert space of Eq.~(\ref{fermionH}) possesses a spectral gap, $\Delta E$, above a degenerate manifold of states for the parameters we consider here \cite{dorier:2005}.   But the entropy gain, $S$, in the free energy cost to create excitations, $\Delta E-TS$, can overwhelm the energy gap depending on the effective dimensionality of excitations.  Strong interactions, $V_{y}>4t$, require the creation of entire domains  (with a perimeter $\sim L$, $\Delta E\sim L$, and $S\sim L$) to destroy non-local correlations as opposed to $\Delta E\sim \mathcal{O}(1)$ and $S\sim \log L$ for $V_{y}<4t$.  Favorable entropy scaling implies that non-local correlation between MF $y$-columns in 2D is much more thermodynamically stable than between pairs of individual MFs in 1D.

\noindent
{\it QMC Test of Thermal Stability:} We test the robustness of SOs of MFs with QMC simulations \cite{sandvik:2002} on Eq.~(\ref{fermionH}) \cite{supplementarymaterial}.  The non-local correlation between edge states at $i=1$ and $i=L$ is captured by a set of $L$ SOs that stretch across each $x$-row:  $P_j\equiv\prod_{i=1}^{L}(1-2n_{i,j})=(-1)^{\sum_in_{i,j}},$
where $j=1,2,\cdots,L$ along $y$.  $P_j$ is equivalent to the fermion parity for the $j$th row.     

The expectation value of the SOs, $P_j$, act as order parameters.  Unique values, $\langle P_{j} \rangle=\pm 1$, can be used to define each sector and therefore indicate stability in the non-local correlations between MFs.  But $\langle P \rangle=0$ indicates that thermal excitations destroy any distinction between sectors.  We compute  $\langle P_{j} \rangle$ to show spontaneous breaking of these discrete symmetries for $V_{y}>4t$ even at non-zero temperatures.  To detect such a symmetry breaking we perturb the above spinless fermion model with a weak global field: $H=H_F-\tilde{h}\sum_{j=1}^{L}P_j.$
The global field, $P=L^{-1}\sum_{j=1}^{L}P_j$,  imposes a splitting between the otherwise degenerate states.   We define $\tilde{h}=hL$ to ensure that the perturbing term imposes a non-zero energy splitting per particle, $h$,  between degenerate sectors even in the limit $L\rightarrow\infty$.  $h> 0$ favors $\langle P \rangle=1$.  

\begin{figure}[t]
\vspace{-0.5in}

\includegraphics[width=4in]{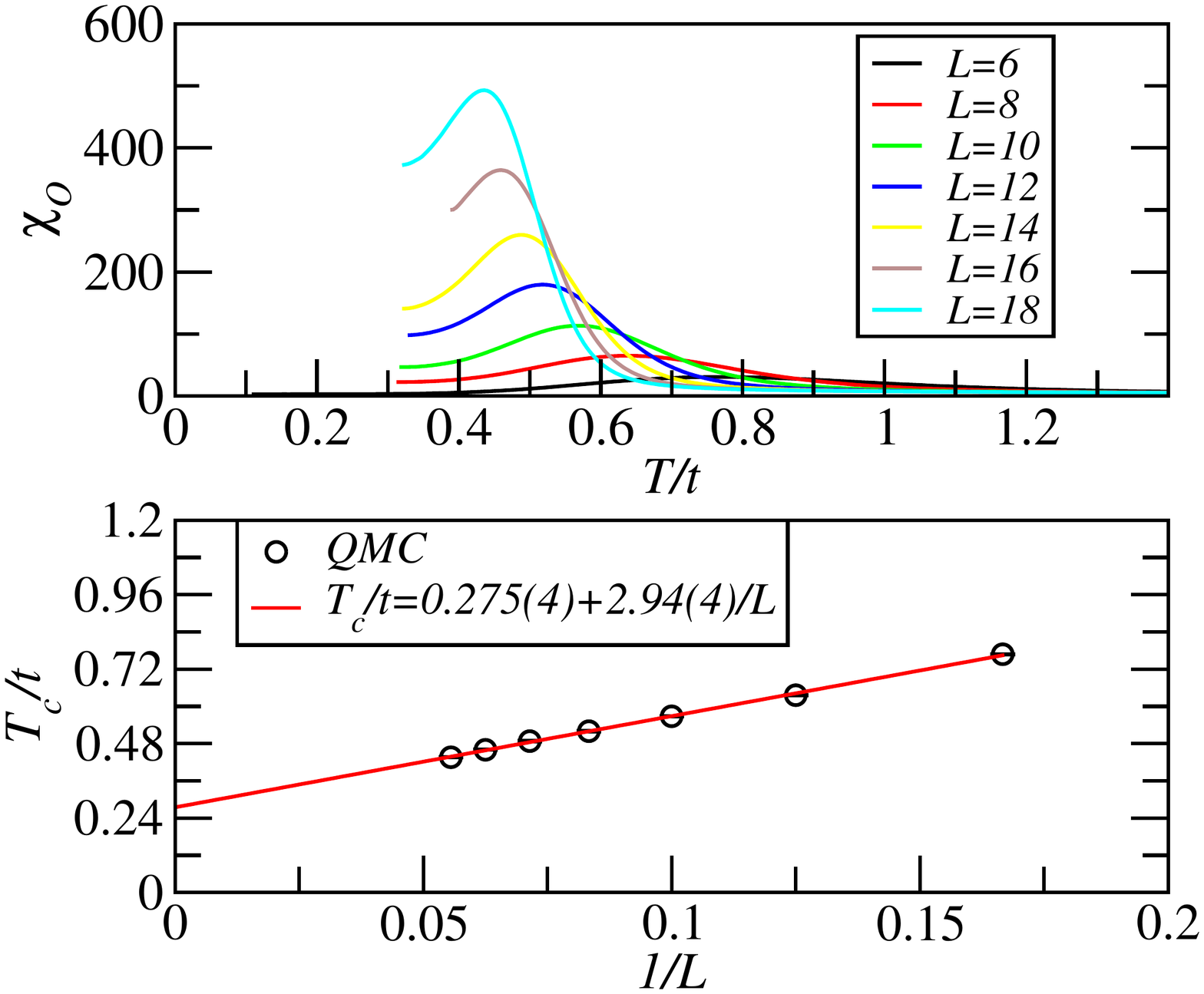}
\vspace{-2.8in}

\includegraphics[width=1.5in]{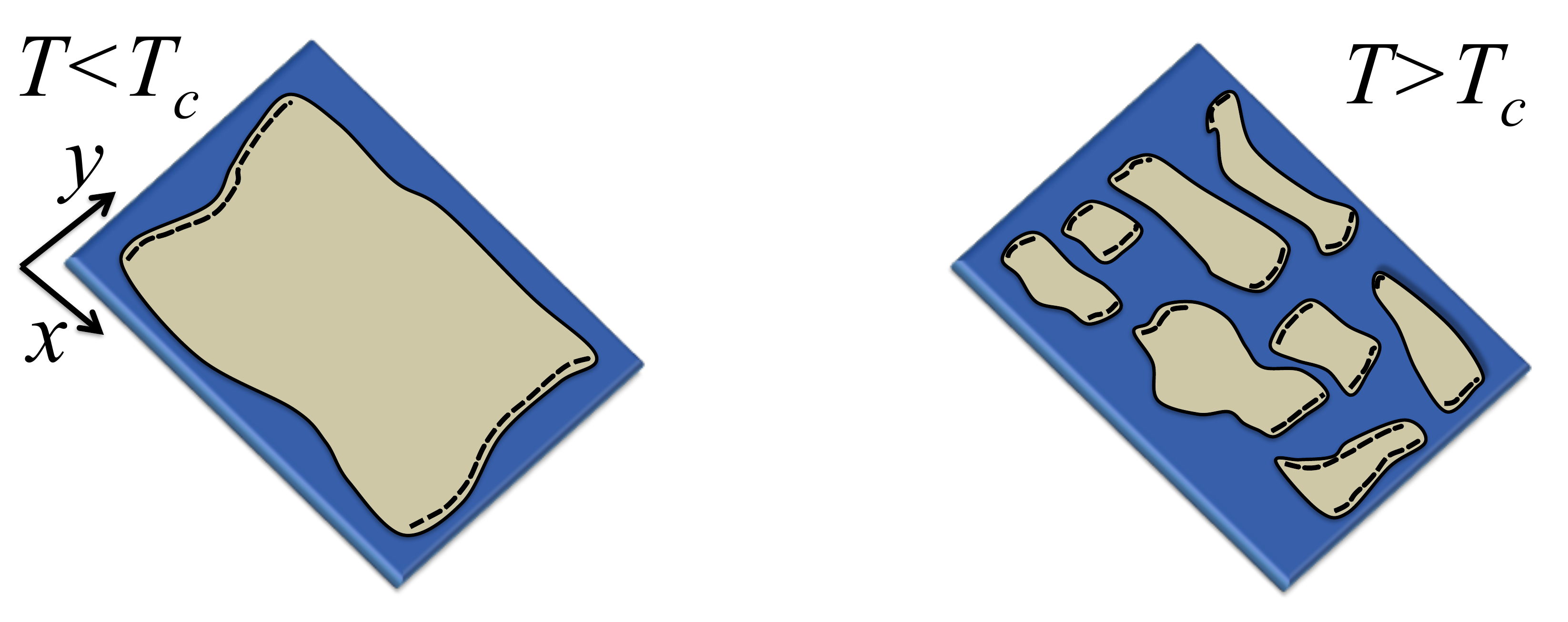}
\hspace{0.7in}
\vspace{2.in}

\caption{(Color online) Top:  The susceptibility of the string-string correlation function $O$ from QMC simulations for different $L$'s at $V_y=4.8t$ and $\mu=0$. The SOs tend to order along the $y$ direction for $T<T_{c}$.  The inset shows a schematic of an ordered domain with MFs forming columns at the ends (dashed lines).  The domains shrink for $T>T_{c}$.  Bottom: $T_{c}$ extrapolated to $L\rightarrow\infty$.  The solid line is a linear chi-squared fit. }
\label{Tc_vs_L}
\end{figure}

We first compute  $\langle P \rangle$ in the limit $V_{y}<4t$ using QMC.  For $V_{y}=3.2t$ we find $\langle P \rangle \rightarrow 0$ with increasing $L$.  This indicates that the SOs in 1D $x$-rows alone are extremely sensitive to thermal fluctuations, as expected from the entropy argument above, even with $\Phi_{i,j}$ held constant.  Our calculations are time independent.  One may find $|\langle P \rangle|>0$ at short times.

We now calculate $\langle P \rangle$ in the strongly interacting case, $V_{y}=4.8t$, where we expect arrays of strings to form stable domains. Fig.~\ref{PvsZ2H} shows $\langle P \rangle$ at low and high temperatures.  
At high $T$ the bottom panel shows that a large value of $h$ is needed to stabilize the SOs.  But at low $T$ (top panel) we find that very small fields tend to force all $x$-rows to spontaneously occupy the lowest energy state in the limit $h\rightarrow 0$, which indicates that $y$-columns of MFs located at $i=1$ and $i=L$ can be prepared in a long-lasting entangled state stretching over large distances even at finite temperatures.

\noindent
{\it Thermal Stability of Domains:}  The arrays of SOs defining domains are stable at low temperatures but eventually break up at large $T$.  To find the critical temperature for domain formation, we define a string-string order parameter that captures the ordering strength along the $y$ direction: $\langle O \rangle\equiv L^{-2}\sum_{j,j'=1}^{L} \langle P_j P_{j'} \rangle.$
The operator $O$ is similar to the static structure factor, $S_{k_{y}}\propto\sum_{j,j'=1}^{L} \exp{[-\bold{i} k_{y}(j-j')]} \langle n_j n_{j'} \rangle$, but with the replacement  $n_j n_{j'}\rightarrow P_j P_{j'}$ and with wavevector $k_y=0$. 

 We look for long-range order in the susceptibility of $O$, $\chi_{O}=L^2(\langle O^2\rangle-\langle O\rangle^2)/T$.  A peak in $\chi_{O}$ versus $T$ indicates the critical temperature $T_c$ at which the large domain breaks up along the $y$ direction.  For $V_{y}<4t$ we find no peaks in our simulations and therefore no domain formation for weakly interacting chains, i.e., $T_{c}=0$.

We observe domain formation in $\chi_{O}$ for $V_{y}>4t$.   The top panel of Fig.~\ref{Tc_vs_L} shows $\chi_{O}$ as a function of temperature for $V_{y}=4.8t$.    Above $T_{c}$ the $y$-columns of MFs are no longer ordered.  The bottom panel extracts $T_{c}$ in the thermodynamic limit, yielding $T_c=0.275(4)t$.  Our results agree with studies on the quantum compass model showing a 
thermal phase transition in the universality class of the 2D Ising model \cite{wenzel:2008}.  

\begin{figure}[t]
\includegraphics[width=3.6in]{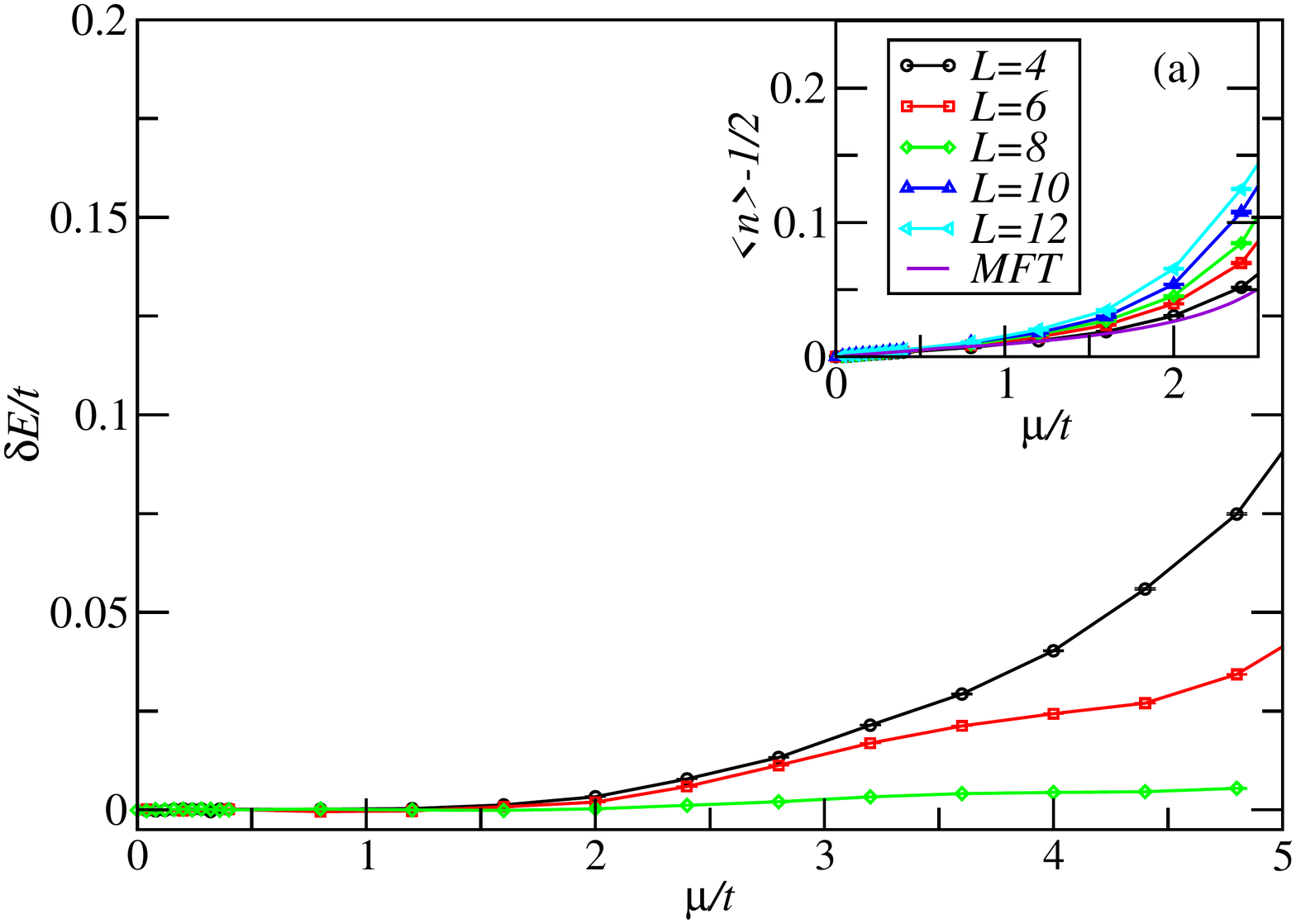}
\vspace{-2.5in}

\vspace{0.63in}
\includegraphics[width=2.03in]{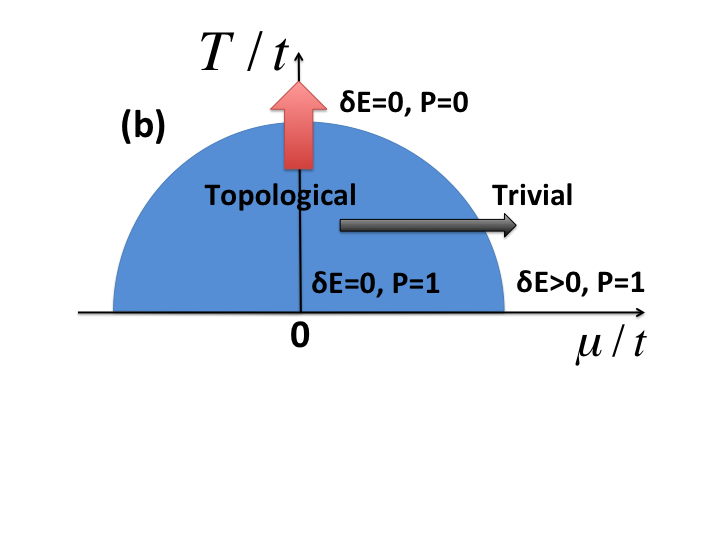}
\hspace{0.7in}
\vspace{-0.1in}

\caption{(Color online) The main panel plots the energy splitting between two sectors defined by $P_{j}=\pm1$ for all $x$-rows as a function of chemical potential for Eq.~(\ref{fermionH}) at $T=0.16t$ and $V_y=4.8t$.  Inset (a) shows a weak linear increase in density with increasing $\mu$ inside the topological phase ($\mu\lesssim 1.5t$).  Inset (b) shows a schematic phase diagram established by the lifting of the degeneracy, horizontal arrow.  The vertical arrow indicates the thermal phase transition explored in Fig.~\ref{Tc_vs_L}. MFT denotes the mean field theory result.}
\label{ESplitting}
\end{figure}

The robustness of the ground state degeneracy also reveals the stability of the SOs.  We denote each ground state energy sector by $E(P_1,P_2,\cdots)$. We found that this degeneracy was not lifted with a weak staggered chemical potential, inter-chain hopping, or a uniform chemical potential shift \cite{lin:2013}.  We present representative results for the uniform chemical potential shift.  Fig.~\ref{ESplitting} shows the energy splitting per particle of two different sectors of the $P_j$ operator: 
$\delta E\equiv E(-1,-1,\cdots)-E(1,1,\cdots)$, as a function of $\mu$.  The flat portion for $\mu/t\ll1$ indicates a robust degeneracy.  Above $\mu\approx 1.5t$ the energy splitting acquires a size dependence, as expected for $\mu>\Delta E$. Inset (a) shows that the particle density has weak linear dependence for $\mu/t\ll1$ which is also captured by the mean field theory.  Our results are consistent with the formation of a thermally robust topological phase, shown in inset (b) of Fig.~\ref{ESplitting}.  

\noindent
{\it Detection in Optical Lattices:}  Domain formation can be observed directly in time-of-flight measurements.  Noise correlations between shots of individual time-of-flight images relate to $S_{k}$ \cite{altman:2004}.  In the topological phase we anticipate the formation of lines, rather than peaks, in noise correlations because the $V_{y}$ term correlates the density along just the $y$ direction for $T<T_{c}$.  Observations of these lines should therefore allow identification of $T_{c}$.

Correlation between MFs could be demonstrated through non-local measures similar to those  proposed in quantum wires \cite{tewari:2008}.  Local spectroscopic probes \cite{jiang:2011,kraus:2012} applied at each domain edge could be adapted to detect the response of one domain edge when dipoles are added to alternating Kitaev chains on the opposite edge.  The particle number parity in the opposite edge should respond with signatures of non-local correlations in dynamics \cite{tewari:2008}.  Recent experiments using high resolution spectroscopy to measure particle number parity \cite{simon:2011} and SOs \cite{endres:2011} could be used to explicitly measure response.

\noindent
{\it Fluctuations in Pairing:}  We connected a model of oriented fermionic dipoles,  Eq.~(\ref{dipolarH}), to a pairing model, Eq.~(\ref{fermionH}).  The pairing model itself demonstrates significantly enhanced stability of MF state via domain formation at $T>0$.  But our specific implementation still allows fluctuations of the pairing field between $x$-rows.  Fortunately, the long-range dipolar interaction has been found to enhance the stability of $p$-wave superfluidity \cite{liu:2012}. 

Coherent reservoirs can further suppress pairing field fluctuations via the proximity effect \cite{diehl:2011,kraus:2012,fu:2008}.  We can show that an optical lattice geometry allowing proximity coupling is possible \cite{supplementarymaterial}.  We note, however, that excitations in the system may couple to those in the reservoir \cite{fidkowski:2011}.

\noindent
{\it Conclusion:}  We considered an effective model of oriented dipolar fermions in a 2D lattice that allows hopping along directions where the dipoles attract but suppresses hopping along directions where dipoles repel.   In the $p$-wave superfluid regime we model the system with  repulsive Kitaev chains.   Each chain experiences a self-consistently renormalized chemical potential due to its neighbor to impose an energy penalty for excitations.  This energy penalty is the mechanism behind MF domain formation and therefore enhances correlation between columns of MFs along each domain edge.  Unbiased QMC confirms that string operators defining non-local MF states remain robust to thermal fluctuations.  

Our approach generalizes to a variety of lattice geometries and even other models with MFs provided they take a similar form: 
$
\sum_{a}H_{\text{M}}^{a}+\sum_{a,b}V_{\text{int}}^{a,b},
$
where $H_{\text{M}}^{a}$ defines a model with MFs, $V_{\text{Int}}^{a,b}$ creates domains with diagonal interactions between models $a$ and $b$, and $V_{\text{Int}}^{a,b}$ does not commute with $H_{\text{M}}^{a}$ \cite{hastings:2011}.  This class of Hamiltonians also applies to Coulomb-coupling in MF models of quantum wire arrays or quasi-1D tubes containing topological superconductors.  

We thank R. Lutchyn, S. Tewari, M. Troyer, and C. Zhang for helpful discussions.  We acknowledge support from the ARO (W911NF-12-1-0335), AFOSR (FA9550-11-1-0313), and DARPA-YFA (N66001-11-1-4122), and computation time at the Lonestar cluster in the Texas Advanced Computing Center.  During preparation of this manuscript we became aware of work on similar non-local order parameters \cite{bhari:2013}.

\section{Supplementary Material for ``Enhancing the thermal stability of Majorana fermions with redundancy using dipoles in optical lattices''}

\subsection{Derivation of Effective Model}
In this section we derive the effective model $H_F$ [Eq.~(2) in the main text] from the dipolar model $H_D$ [Eq.~(1) in the main text] at the Hartree-Fock level.  This shows that, deep in the superfluid phase, $H_{F}$ captures the essential physics of $H_{D}$.  All of our numerical 
calculations in the paper are performed on $H_F$.      

The attractive interaction term along $x$-rows, $V_x(\theta)n_{i,j}n_{i+1,j}$, in $H_D$ decouples in the Hartree-Fock approximation:
\begin{eqnarray}
n_{i,j}n_{i+1,j}
&\approx&\langle n_{i,j}\rangle n_{i+1,j}+\langle n_{i+1,j}\rangle n_{i,j} \nonumber \\
&-&\langle a_{i,j}^{\dagger}a_{i+1,j}\rangle a_{i+1,j}^{\dagger}a_{i,j}+\langle a_{i,j}^{\dagger}a_{i+1,j}^{\dagger}\rangle a_{i+1,j}a_{i,j}\nonumber\\
&-&C + h.c.,
\end{eqnarray}
where $C\equiv\langle n_{i,j}\rangle\langle n_{i+1,j}\rangle-\langle a_{i,j}^{\dagger}a_{i+1,j}\rangle\langle a_{i+1,j}^{\dagger}a_{i,j}\rangle+\langle a_{i,j}^{\dagger}a_{i+1,j}^{\dagger}
\rangle\langle a_{i,j}a_{i+1,j}\rangle$. We define the renormalized chemical potential $\mu=\mu_0+2\langle n_{i,j}\rangle|V_x(\theta)|-V_y/2$ and the renormalized hopping 
$t=t_x-|V_x(\theta)|\langle a_{i+1,j}^{\dagger}a_{i,j}\rangle$. We further assume that by tuning $V_x(\theta)$ the renormalized hopping $t$ matches the pairing amplitude  
$t=|V_x(\theta)|\langle a_{i+1,j}^{\dagger}a_{i,j}^{\dagger}\rangle$.
As argued in the main text, we also take the $t_y=0$ limit to arrive at the effective model $H_F$ in Eq.~(2) of the main text.

\subsection{Ground State Degeneracy}
In this section we show that the ground state of $H_F$ in the main text is $2^L$ fold degenerate for our cylindrical geometry \cite{doucotsupp:2005, doriersupp:2005} for $t_{y}=0$.  We then discuss the $t_{y}\rightarrow 0$ limit. In the main text we defined a set of SOs $P_j$ along the $x$ direction, which commute with $H_F$. Similarly, we define a set of SOs $Q_i$ along the $y$ axis, which also commute with $H_F$,
\begin{equation}
Q_i=\prod_j(2\tilde{a}_{i,j}),
\end{equation}
where $i=1,2,\cdots,L$ and 
\begin{equation}
\tilde{a}_{i,j}\equiv F_{i,j}(a_{i,j}^{\dagger}+a_{i,j}^{\phantom{\dagger}})/2,
\label{Sx_definition}
\end{equation} 
where the transformation coefficients are given by:
\begin{equation}
F_{i,j}=\prod_{j'<j}\prod_k(1-2n_{k,j'})\prod_{i'<i}(1-2n_{i',j}).
\end{equation}
Note that the operator 
$\tilde{a}_{i,j}$ corresponds to a spin $\frac{1}{2}$ operator along the $x$ direction in spin space, $S_{i,j}^x$, based on the Jordan-Wigner transformation \cite{chensupp:2007}.  One can check that $\{P_j, Q_i\}=0$.

To see the degeneracy explicitly, suppose that we have a common eigenstate $\phi_0$ of $H_F$ and $Q_i$. If we act $P_j$ on the state $\phi_0$, we get $\phi_1=P_j\phi_0$. Since $P_j$ does not commute with $Q_i$, $\phi_1$ must be different from $\phi_0$. However, $\phi_1$ is still an eigenstate of $H_F$ with the same eigenvalue as $\phi_0$, because $P_j$ commutes with $H_F$. Each eigenstate is, therefore, at least 2-fold degenerate. Furthermore, since $[P_kP_j,Q_i]=0$, $\phi_1$ is also an eigenstate of the operator product $P_kP_j$. We then have $\phi_1=P_j\phi_0 \propto (P_kP_j)P_j\phi_0=P_k\phi_0$, which means that acting $P_k (k\neq j)$ on $\phi_0$ will not generate a different state than $\phi_1=P_j\phi_0$. Every eigenstate, including the ground state, is therefore, 2-fold degenerate. 

Exact diagonalization studies in combination with $L^{\text{th}}$ order perturbation theory show that in the $L\rightarrow\infty$ limit the low-lying $2^L-2$ excited states will collapse with the exact 2-fold degenerate ground state, thus forming a $2^L$-fold degenerate ground state in the equivalent spin-quantum compass model  \cite{doriersupp:2005} (For a mapping to the quantum compass model see the section ``QMC Simulations'').  The gap between the ground state and the low-lying $2^L-2$ excited states was found to collapse as $\sim (2t_{x}/V_{y})^{L}$ for $V_{y}>4t_{x}$ \cite{doriersupp:2005}. Note that the $2^L$-fold degeneracy arises even in the large $V_y$ limit.

We now consider the $t_y\rightarrow 0$ limit, i.e., non-zero hopping along the $y$ direction.  In our model, with $t_y=0$, edge MFs are unable to hybridize with those in neighboring rows. In the $t_y\rightarrow 0$ limit we also observe a $2^L$ degeneracy in spite of edge MF coupling (hybridization) effects discussed in the literature \cite{pottersupp:2010}. Our model is different from these works because it is very strongly interacting.  Even with a small $t_y$ hopping, we believe that hybridization is still strongly suppressed because of the strong $V_y$ term, which will give a large energy penalty if a single fermion hops between chains. We have performed direct numerical simulations of Eq.(1) in the main text for various lattice sizes, $L=4,6,$ and 8, to confirm, within numerical accuracy, the emergence of such a set of degeneracies in the ground state.  For example, we find degeneracies 
for $t_x=1$, $V_y=1.2$, and $V_x=-0.053$, that are immune to small $t_y$ perturbations.

\subsection{Validating a Mean Field Picture}

To show the existence of MFs and domains we perform a mean field decoupling of Eq.~(2) in the main text along the $y$ direction.   The mean field theory presented in this section is in terms of real fermions but is equivalent to the MF mean field theory presented in the next section, Eq.~(\ref{MFmajoranaH}), and in the main text.  We then verify the mean field theory by direct comparison with an unbiased QMC analysis. Finally we will discuss the parameter regimes of validity. 

To construct the mean field equations we divide the lattice into 2 sublattices, $A$ and $B$, along the $y$ direction, and decouple the interaction terms (staggered density 
assumption). We obtain the following 4 coupled mean field equations:  
\begin{eqnarray}
H_1^{\alpha}&=&-8t\langle \tilde{a}_{i+1,\alpha}\rangle\tilde{a}_{i,\alpha}-\tilde{\mu}_{\alpha}\left (n_{i,\alpha}-\frac{1}{2} \right ),\nonumber\\
H_2^{\alpha}&=&-t\sum_i \left(a_{i,\alpha}^{\dagger}-a_{i,\alpha}^{\phantom{\dagger}}\right ) \left (a_{i+1,\alpha}^{\dagger}+a_{i+1,\alpha}^{\phantom{\dagger}} \right)\nonumber\\
     & & -\tilde{\mu}_{\alpha}\sum_in_{i,\alpha},
\label{mftH}
\end{eqnarray} 
where $\tilde{\mu}_{\alpha}=\mu_{\alpha}-2V_y\langle n_{i,\alpha}-1/2\rangle$.  
$\mu_A$ and $\mu_B$ are applied staggered chemical potentials for $A$ and $B$ sublattices.  
 In the spin language, the first equation defines a single spin in a magnetic field while the second is a quantum Ising model.  We use the solutions of both of these models \cite{lieb:1961,pfeuty:1970} to solve both models exactly and then the coupled equations, Eqs.~(\ref{mftH}), through iteration.

\begin{figure}[t]
\includegraphics[clip,width=\columnwidth]{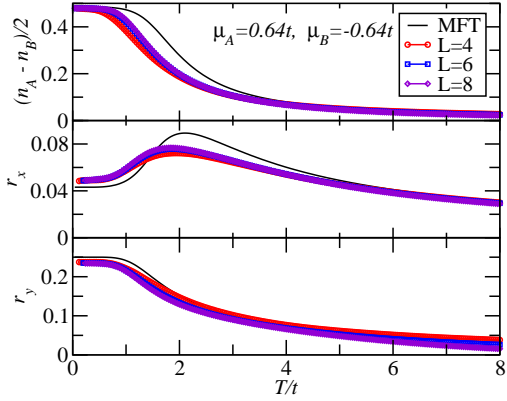}
\caption{ (Color online.) QMC (L=4, 6, 8) and mean field theory comparison of the staggered density (top), intra-$x$-row hopping and pairing correlation function (middle), and the inter-$x$-row density-density correlation function (bottom) at $V_y=4.8t$. We apply staggered chemical potentials $\mu_A$ and $\mu_B$ to the $A$ and $B$ sublattices, respectively. }
\label{QMC_PMFT_cmp}
\end{figure}

Eqs.~(\ref{mftH}) assume a spatially uniform chemical potential (for each sublattice).  If this assumption is correct, it implies that excitations for any given $x$-row are copied to all other $x$ rows to yield a domain.  The existence of domains of string operators is therefore implicit in the mean field theory but we must validate Eq.~(\ref{mftH}) as a good approximation to Eq.~(2) in the main text to justify this picture.

We validate Eqs.~(\ref{mftH}) by direct comparison with QMC solutions to Eq.~(2) in the main text. To compare we compute correlation functions using both mean field theory and QMC.  The following local correlation functions define quantum bond order along the $x$ direction and density bond order along the $y$ direction.
\begin{eqnarray}
r_x&\equiv&\frac{1}{4}\langle (a_{i,j}^{\dagger}-a_{i,j}^{\phantom{\dagger}})(a_{i+1,j}^{\dagger}+a_{i+1,j}^{\phantom{\dagger}})\rangle,\nonumber\\
r_y&\equiv&\langle (\frac{1}{2}-n_{i,j})(n_{i,j+1}-\frac{1}{2})\rangle.
\label{rdefinitions}
\end{eqnarray}
Under the spin mapping these correlation functions have been studied in a corresponding spin model, the quantum compass model \cite{nussinov:2005,scarola:2009}.  

Fig.~\ref{QMC_PMFT_cmp} shows that the mean field theory offers an excellent approximation to the QMC results.  The large value of $V_{y}$ leads to bond ordering along $y$ (large $r_{y}$).  But the non-zero values of $r_{x}$ show quantum correlations along the $x$ direction.   Therefore both QMC and mean field theory show that the $y$-columns superpose throughout the lattice to yield a quantum entangled ground state at non-zero temperatures.  The good agreement between QMC and mean field theory therefore supports the domain picture implicit in Eqs.~(\ref{mftH}). 

There are, however, small differences between QMC and mean field calculations for $T/t<4$ in Fig.~\ref{QMC_PMFT_cmp}. This is due to the fact that mean field calculations ignore quantum fluctuations (and therefore underestimate $r_x$) at low temperatures and exaggerate the effects of classical $V_y$ interactions (and therefore overestimate $r_y$). Despite this 
drawback, mean field calculations for $V_y>4t$ still capture the essential physics of the original model. To be specific, at low temperatures both QMC and mean field calculations give $r_y=1/4$, 
which means that $(\langle n_{i,j}\rangle+\langle n_{i,j+1}\rangle)/2-\langle n_{i,j}n_{i,j+1}\rangle=1/2$.  At half filling for a uniform system, i.e., $\langle n_{i,j}\rangle+\langle n_{i,j+1}\rangle=1$, we have $\langle n_{i,j}n_{i,j+1}\rangle=0$, which shows that the system avoids large $V_y$ interactions. This 
explains why mean field calculations are accurate in this regime.  

The validity of our mean field theory crucially depends on the order parameter assumption (staggered density in a given column to avoid $V_y$ interactions).  Mean field theory breaks down when different ordering appears. 
This is shown in inset (a) of Fig.~4 in the main text for the large $\mu/t$ limit.  Here the topological phase disappears. 
In this limit a new order parameter is required to capture the effects of adding extra particles to the system.

\subsection{Mapping to Majorana Fermions}
Here we prove that we can transform Eq. (2) in the main text into an interacting MF model by introducing two MF operators, $c_{2i,j}$ and $c_{2i-1,j}$, for each site of the lattice, $(i,j)$ \cite{kitaevsupp:2001} with $c^{\vphantom{\dagger}}_{a,b}c^{\vphantom{\dagger}}_{a',b'}=-c^{\vphantom{\dagger}}_{a',b'}c^{\vphantom{\dagger}}_{a,b}$ (for $\{a,b \}\neq\{a',b' \}$), $c^{\vphantom{\dagger}}_{a,b}=c_{a,b}^{\dagger}$ and $(c^{\vphantom{\dagger}}_{a,b})^2=1$.  The absence of kinetics along the $y$ direction implies that each particle can be labeled with a specific $x$-row index, $j$.   The MF operators then relate to the physical fermion operators by a complex superposition:
$
a_{i,j}^{\dagger}=(c_{2i-1,j}-\bold{i}c_{2i,j})/2.
$
We can now demonstrate the existence of edge states by mapping Eq.~(2) in the main text to MF space:
\begin{eqnarray}
H_M=& &\bold{i}t\sum_{i,j}c_{2i,j}c_{2i+1,j} +\frac{\bold{i}\mu}{2}\sum_{i,j}c_{2i-1,j}c_{2i,j}\nonumber\\
&-&\frac{V_y}{4}\sum_{i,j}c_{2i-1,j}c_{2i,j}c_{2i-1,j+1}c_{2i,j+1}.
\label{majoranaH}
\end{eqnarray}
Here we see that the first two terms equate to the Kitaev chains [the first term $\sum_j H_K^j$ in Eq.~(2) in the main text] and define a bilinear MF theory.  States defined by the dangling operators, $c_{1,j}$ and $c_{2L,j}$, at the ends of each $x$-row establish two-fold degenerate MF states that can be entangled at $T=0$.

Next we want to understand the effect of interactions, $V_y>0$, on the degenerate MF states in a mean field approximation (validated in the main text and in the previous section). We note that the MF correlation function is directly related to the real fermion number operator: $C^M_{i,j}\equiv (\bold{i}/2)c_{2i-1,j}c_{2i,j}=n_{i,j}-1/2$. From the mean field and QMC comparison result and discussions in the previous Supplementary Material section [see $r_y$ in Eq.~(\ref{rdefinitions}) and Fig.~\ref{QMC_PMFT_cmp}], we can see that at low temperatures for fixed index $i$ the MF correlation function $C^M_{i,j}$ has alternating values of $\frac{1}{2}$ and $-\frac{1}{2}$ along the $y$ direction.  This minimizes the interaction energy. Therefore, we can do a mean field decoupling of the $V_y$ interaction term in the MF Hamiltonian, Eq.~(\ref{majoranaH}), to obtain the 
following Hamiltonian:
\begin{equation}
H_M^{\alpha}=\bold{i}t\sum_{i=1}^{L-1}c_{2i,\alpha}c_{2i+1,\alpha} +\frac{\bold{i}\tilde{\mu}_{\alpha}}{2}\sum_{i=1}^{L}c_{2i-1,\alpha}c_{2i,\alpha},
\label{MFmajoranaH}
\end{equation}
where $\alpha \in \{ A,B\}$ indexes sublattices and $\tilde{\mu}_{\alpha}=\mu+V_y\langle C^M_{i,\alpha} \rangle$.

Eq.~(\ref{MFmajoranaH}) yields edge MFs only for certain parameter regimes.  To see where, we solve the eigenequation 
$H_M^{\alpha}u_{\alpha}=0$ for the zero-energy eigenfunction $u_{\alpha}$ of the $\alpha$'th Kitaev chain. One real-space solution is \cite{kitaevsupp:2001}:
\begin{equation}
u_{\alpha}\propto \bigg (1,0,\frac{\tilde{\mu}_{\alpha}}{2t},0,\left (\frac{\tilde{\mu}_{\alpha}}{2t}\right )^2,0,\cdots \bigg).
\end{equation}
Here we see that the edge MF survives for $\tilde{\mu}_{\alpha}/2t\ll 1$.  At half filling ($\mu=0$) this gives highly localized edge MFs, $u_{\alpha}\propto (1,0,0...)$.  For $V_y>4t$ $C^M_{i,j}$ oscillates in sign for a single classical configuration but gives $\langle C^M_{i,j}\rangle=0$ in the quantum ground state.  This shows that $\tilde{\mu}_{\alpha}=\mu$, i.e., the chemical potential for each Kitaev chain is not renormalized for $V_y>4t$.  But the situation is different for 
$V_y<4t$.  Here we have $\tilde{\mu}_{\alpha}\sim \mu+V_y$.  In this regime, the large chemical potential prevents the formation of edge MFs.

\subsection{QMC Simulations}

In this section we describe our QMC simulations in more detail.  We first show that, after mapping Eq.~(2) in the main text to a spin model, we can compute correlation functions using the Stochastic Series Expansion (SSE) \cite{sandviksupp:2002} combined with the quantum 
Wang-Landau (QWL) algorithm \cite{troyersupp:2003}.  QMC parameters are given.  We then discuss the nature of the sign problem that arises when we add inter-chain tunneling to simulate Eq.~(1) in the main text.  

We first show how to map Eq.~(2) in the main text to a spin model.  We use a Jordan-Wigner transformation that zig-zags through the lattice \cite{chensupp:2007}:
\begin{eqnarray}
a_{i,j}&=&\bigg( \prod_{i'<i,j'}\sigma_{i',j'}^{z}\prod_{j''=1}^{j-1}\sigma_{i,j''}^{z} \bigg)\sigma_{i,j}^{+}, \nonumber\\
\sigma_{i,j}^{z}&=&(-1)^{a_{i,j}^{\dagger}a_{i,j}^{\vphantom{\dagger}}},
\end{eqnarray}
where $\sigma^x$, $\sigma^y$, and $\sigma^z$ are the Pauli matrices and $\sigma^{\pm}=(\sigma^x\pm\bold{i}\sigma^y)/2$, 
 to map the model onto the quantum compass model \cite{chensupp:2007}:
 \begin{eqnarray}
 H_F=\sum_{i,j}\left [-t\sigma_{i,j}^{x}\sigma_{i+1,j}^{x}+\frac{V_y}{4}\sigma_{i,j}^z\sigma_{i,j+1}^z 
 -\mu_{0}\frac{1-\sigma_{i,j}^{z}}{2}\right ]\nonumber
 \end{eqnarray}
To solve this model we perform QMC simulations with SSE \cite{sandviksupp:2002} combined with the QWL algorithm \cite{troyersupp:2003}. 

In the QWL approach the partition function is expanded as a series in powers of $\beta\equiv (k_{B}T)^{-1}$:
\begin{equation}
\textrm{Tr} e^{-\beta H_F}=\sum_{n=0}^{N_{\textrm{max}}}S \vert g(n)\vert \beta^n,
\end{equation}
where $N_{\textrm{max}}$ is the maximum expansion order.  $N_{\textrm{max}}$ determines the lowest temperature that can be reached in the simulation and
$g(n)$ corresponds to the classical density of states. $S$ is the overall sign.  In the absence of a sign problem we have $S=1$ and $g(n)=\vert g(n)\vert$.  In the presence of a sign problem we have $\langle S\rangle <1$.  Severe sign problems, $\langle S\rangle \rightarrow 0$, prevent control of error in QMC sampling.  
The quantum compass model does not have a sign problem, implying that Eq.~(2) in the main text does not have a sign problem.

The distribution of $g(n)$ is obtained from a random sampling protocol \cite{troyersupp:2003}. It can be used to 
estimate the free energy, internal energy, entropy, heat capacity, and other properties of the system. We note that to measure other physical quantities, e.g., the density, density-density correlation, and the fermion parity operator, we need to accumulate their distributions at every order of the series expansion.

In simulating $H_{F}$ we find that the energy barrier between different fermion parity operator sectors is very large. The large energy barrier dramatically increases the autocorrelation time in conventional QMC simulations with non-local updating.  Without the QWL algorithm, the energy autocorrelation time for $V_y>4t$ is typically $\sim 10^3 - 10^4$ MC sweeps, which is prohibitively large for obtaining accurate QMC results.  (We define 1 MC sweep as 1 diagonal update followed by $ N_{\rm max}/L_{\rm loop}$ loop updates with average loop length $L_{\rm loop}$.)  We find that the QWL algorithm is necessary to reduce the autocorrelation time in QMC by enabling tunneling between different fermion parity sectors.

We check the convergence of various physical quantities in the simulation with respect to $N_{\rm{max}}$.  We find that local quantities such as internal energy, average density, density-density correlation function, etc., converge much faster than the non-local fermion parity operator, $P$, at low temperatures, which usually requires a much larger $N_{\rm{max}}$. In practice we find the following values for $N_{\rm{max}}$ to be enough for $P$ to converge in our simulations in the desired low temperature range:
$N_{\rm{max}}=5000, 8000,$ and $10000$ for $L=4, 6$, and $8$, respectively. A typical QMC run on a single 2.53 GHz Intel Xeon CPU with the above $N_{\rm max}$ takes 1, 2, and 
12 days, respectively, for the flat histogram to converge within $10^{-6}$. We usually do 10 such runs to estimate the error bars of various physical quantities for each set of parameters.

We now discuss simulation of Eq.~(1) in the main text.  We map into a quantum spin model using the same Jordan-Wigner transformation \cite{chensupp:2007}:
\begin{eqnarray}
H_{QS}&=&\sum_{i,j}\bigg\{-t_x\sigma_{i,j}^{-}\sigma_{i+1,j}^{+}-t_y(-1)^{n_{d}(i,j;i,j+1)}\sigma_{i,j}^{-}\sigma_{i,j+1}^{+}\nonumber\\
& +&h.c.+\frac{V_x(\theta)}{4}\sigma_{i,j}^z\sigma_{i+1,j}^z+\frac{V_y}{4}\sigma_{i,j}^z\sigma_{i,j+1}^z\nonumber\\
&-&\mu_{0}\frac{1-\sigma_{i,j}^{z}}{2}\bigg\},
\end{eqnarray} 
where:
\begin{equation}
n_{d}(i,j;i,j+1)\equiv\sum_{i'=i+1}^L(-1)^{\tilde{n}_{i',j}}+\sum_{i'=1}^{i-1}(-1)^{\tilde{n}_{i',j+1}},
\end{equation}
counts the number of down spins between sites $(i,j)$ and $(i,j+1)$, exclusively. Here $\tilde{n}_{i',j}=1 (0) $ if there is a down (up) spin at site $(i',j)$.  For $t_{y}=0$,  $H_{QS}$ reduces to the quantum compass model discussed above (and therefore Eq.~(2) in the main text).  But the $t_y$ term introduces a sign problem in QMC simulations. 

Despite the sign problem, the above quantum spin model can also be simulated with SSE
combined with the QWL algorithm.  We find that, for small $t_y$, the sign problem is not severe.  For example, for an $L=4$ system and $t_y=t_x/10$, we find $\langle S \rangle >0.2$ for $T>t_{y}$. For smaller 
$t_y$ values, we can approach lower temperatures.  We have performed QMC simulations on the quantum spin model for $L=4, 6$, and $8$ to detect the emergence of the ground state degeneracy.  We discuss an 
example result in the section, ``Ground State Degeneracy''.

\subsection{System-Reservoir Optical Lattice Geometry}

We show that an optical superlattice can be used to host a 2D ``system'' lattice parallel to a 2D ``reservoir'' lattice.  The system lattice is an array of chains in the $x-y$ plane that allow strong tunneling along the $x$-direction and weak tunneling along the $y$-direction.  The reservoir lattice is a square lattice with nearly equal tunneling along both the $x$ and $y$ direction.  The increased dimensionality of the reservoir strengthens the pair superfluid in the reservoir.  A tunable potential barrier controls the tunneling between the system and the reservoir.

\begin{figure}[t]
\includegraphics[clip,width=\columnwidth]{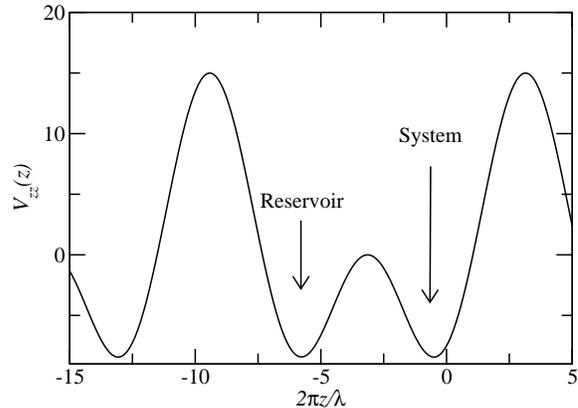}
\caption{Plot of the potential defining a double well optical lattice along the $z$ direction for $v_{z}=-15E_{R}$, $\phi_{1}=0$, and 
$\phi_{2}=3\pi/2$. }
\label{double_well}
\end{figure}

The optical lattice is formed from three laser beam pairs:  1) a double well optical lattice potential, $V_{zz}$, formed from the interference of counter propagating beams along the $z$ direction, 2) a pair of beams with the same polarization counter-propagating in the $x$-$z$ plane, to form $V_{xz}$, and 3) a similar pair of beams but in the $y$-$z$ plane, to form $V_{yz}$.  If each beam pair does not interfere then the total potential experienced by the particles is: $V_{\text{tot}}(x,y,z)=V_{zz}(z)+V_{xz}(x,z)+V_{yz}(y,z)$.  

\begin{figure}[t]
\includegraphics[clip,width=\columnwidth]{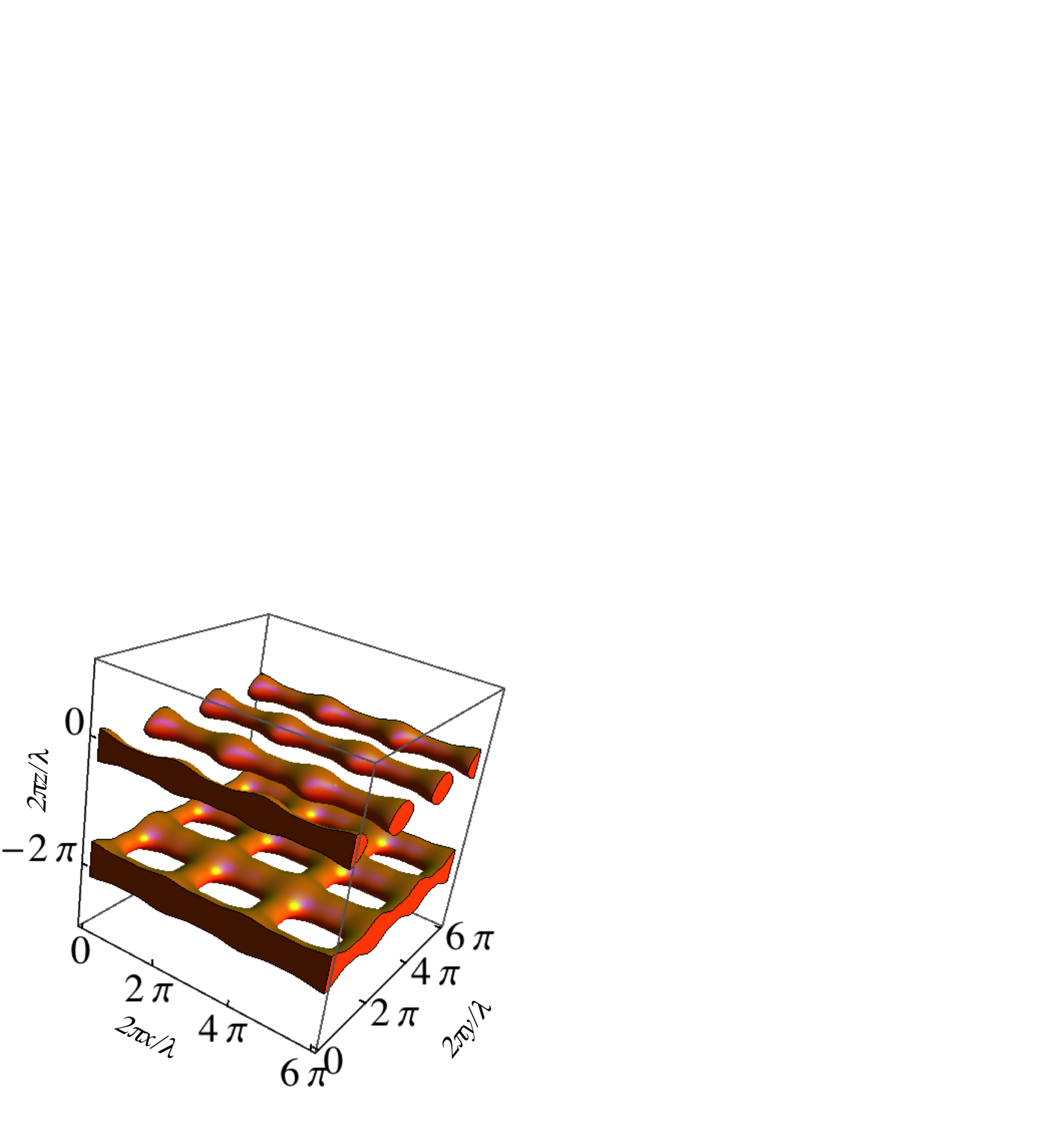}
\caption{(Color online.) Plot of the total potential for the system-reservoir optical lattice, $V_{\text{tot}}$.  Points are plotted for $v_{\text{tot}}< -10 E_{R}$.  The parameters are chosen to be: 
$v_{z}=-15E_{R}$, 
$v_{x}= -0.5 E_{R} $,
$v_{y}= -1 E_{R}$,
$\phi_{1}=-(k\pi+2\pi/1.9)$, and 
$\phi_{2}=-(k\pi/2+2\pi/1.9)$. 
}
\label{bath_lattice}
\end{figure}

The system and reservoir are formed from the double well lattice along the $z$ direction.  The potential $V_{zz}$ can be formed from the interference of two counter propagating lasers with differing wavelengths.  The distance between the system and the reservoir can be changed by using different laser wavelengths to define the double well.  We choose the wavelengths to differ by a factor of 2 to yield:
\begin{eqnarray}
V_{zz}(z)&=&\frac{v_{z}}{2}\left[ \cos\left(kz-\phi_{1}\right)-\cos\left(kz/2-\phi_{2}\right) \right] \nonumber \\
\end{eqnarray}
Here the wavevector of the primary lattice is $k=2\pi/\lambda$. This potential is plotted in Fig.~\ref{double_well}.

We consider an arrangement where the potential established by the remaining beam pairs is given by:
\begin{eqnarray}
V_{xz}(x,z)&=& v_{x} \left[  \cos\left( k x \right)+ \cos\left( k z \right) \right]^{2} \nonumber \\
V_{yz}(y,z)&=& v_{y} \left[  \cos\left( k y \right)+ \cos\left( k z \right) \right]^{2}
\end{eqnarray}
Because the beam pairs forming $V_{xz}$ and  $V_{yz}$ each have the same polarization, they interfere to form a node in the $z$ direction at the location of the reservoir.  The reservoir then experiences a nearly isotropic square lattice even with $v_{x}\neq v_{y}$.

Fig.~\ref{bath_lattice} plots an equipotential surface defined by $V_{\text{tot}}$.  The potentials are defined in units of the lattice recoil, $E_{R}\equiv h^{2}/2m\lambda^{2}$.  Here $m$ is the mass of the particles.  Fig.~\ref{bath_lattice} shows a configuration where the particles in the system lattice, near $z=0$, have little tunneling along $y$ whereas the reservoir lattice, near $z=-\lambda$, is essentially a 2D square lattice.  This geometry allows a 2D dipolar superfluid in the reservoir to be placed in close proximity to the system lattice.

\end{document}